# Influence of intensity distribution of laser beam on the properties of nanoparticles obtained by laser ablation of solids in liquids

P.G. Kuzmin and G.A. Shafeev*

Wave Research Center of A.M. Prokhorov General Physics Institute of the Russian Academy of Sciences, 38, Vavilov street, 119991 Moscow Russian Federation

**Abstract**

Control over the properties of nanoparticles obtained by laser ablation in liquids is experimentally demonstrated via modulation of the beam intensity profile on the target. Mask projection scheme was used with either a copper laser (wavelength of 510. 6 nm, pulse duration of 10ns) or with a Ti:sapphire laser (wavelength of 800 nm and pulse duration of 200 fs). Si and ZnSe were chosen as target materials. Obtained nanoparticles were characterized using Transmission Electron Microscopy (TEM), optical absorption spectroscopy and photoluminescence. It was shown that that size of Si nanoparticles depends on the spatial profile of the laser beam.

*Corresponding author, e-mail shafeev@kapella.gpi.ru

1. Introduction

Laser ablation of solid target in liquids is simple and reliable method for generation of nanoparticles (NPs) of almost any metals and semiconductors. It is an alternative to chemical synthesis of NPs and is capable of producing NPs that are free of any surface-active substances and counter-ions. The properties of NPs generated by laser ablation in liquids depend on a number of experimental parameters. The specific feature of laser ablation in liquids is subsequent interaction of generated NPs with the laser beam via direct absorption of laser radiation. So the final properties of NPs depend on such parameters as laser pulse duration, laser wavelength, laser fluence on the target, nature of surrounding liquid, number of laser shots, etc [1,2]. The necessary condition of NPs formation is melting of the target surface under the laser beam. Then this melt is dispersed into the liquid surrounding the target by recoil pressure of the liquid vapors during their expansion.

One of the experimental parameters was ignored so far, namely, the spatial profile of the laser beam in the plane of the target. Indeed, it is the laser fluence $j$ on the target that is controlled, since the temperature of the target $T$ is proportional to $j$ for laser pulses longer than the time of electron-phonon relaxation. The beam profile, either Gaussian or a flat-top one are considered as a secondary factor that does not influence the properties of generated NPs. However, recent results on the laser ablation of a gold target with a periodic function of lateral coordinate $x$ in the plane of the target $I=I_0(1+sin(kx))$, where $k=2\pi/\lambda$ stands for the wave vector of intensity profile with spatial period $\lambda$, indicate that the intensity profile of the laser spot can significantly affect the properties of NPs generated by laser ablation in liquid [3]. In particular, Au NPs generated with the interference pattern have elongated shape, related to the elongated shape of the melt bath on the target during the laser pulse.

The present communication deals with the effect of the spatial modulation of the beam intensity on the properties of NPs produced by laser ablation of two semiconductors, namely, Si and ZnSe. The modulation of the laser intensity is achieved using well-known mask projection scheme. Ablation with the mask alters the properties of generated NPs.

2. Experimental

The experiment on nanoparticle generation by laser ablation of solids in liquid was described in detail elsewhere [1]. Si and ZnSe were chosen as targets. A Cu vapor laser was used with wavelength of 510.6 nm (another laser output at 578.2 nm was suppressed using selective absorbers), pulse duration of 10 ns, and the repetition rate of 15 kHz was used to produce ZnSe NPs. As the laser source for preparation of Si NPs a femtosecond Ti:sapphire laser with pulse duration of 200 fs and repetition rate 1kHz was used.

The intensity profile of the laser beam was modified with the mask imaged on the target surface (Fig. 1). The size of holes of the mask was 200 μm, which resulted in the 1 – 2 μm ablated spots on the target as it was confirmed by examination of the surface with an optical microscope.

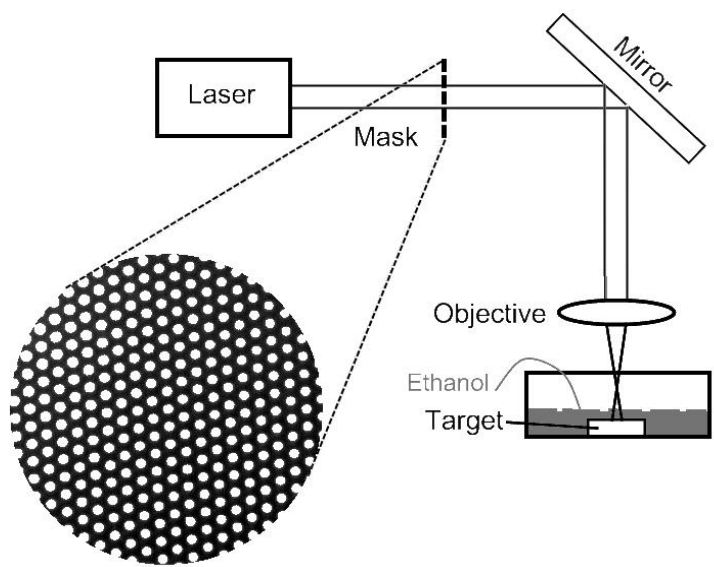

Fig. 1. Mask projection scheme used for generation of NPs via ablation under the layer of a liquid.

Laser radiation was focused onto a target under an ethanol layer from 2 to 3 mm thick. A cell with liquid was placed on a computer-driven stage which moved the target under the laser beam. This precaution was used to avoid large crater formation on the target surface, thus, to retain beam focusing on the target during ablation.

Ethanol (95%) was used as working liquid for ablation. Surfactants were added to liquid after ablation in order to decelerate the coagulation process. For measuring absorption and luminescence spectra of colloidal solutions of nanoparticles produced by ablation an OceanOptics fiber spectrometer in the range of 200 to 850 nm and Andor were used respectively.

The morphology of Si and ZnSe NPs was studied using a Transmission Electron Microscope (TEM). To this end, colloid solutions were evaporated on a copper grid coated with a carbon membrane.

3. Results

3.1 Si nanoparticles

The Si target ablation in ethanol using the radiation of a Ti:sapphire laser results in generation of Si NPs. The absorption spectrum of this solution is shown in Fig. 2, a. An increase in the optical density near 200 nm is explained by ethanol opacity in this spectral region.

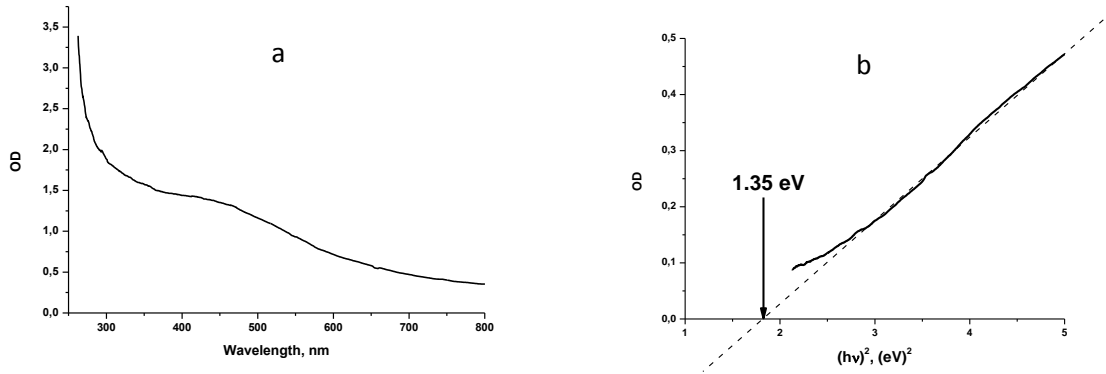

Fig. 2. Optical density of the colloidal solution of Si NPs synthesized by ablation of a bulk Si target with a 200 fs radiation of a Ti:sapphire laser at 800 nm (a). Estimation of the effective bandgap of Si NPs.

Si is an indirect band semiconductor, and the estimation of the effective bandgap of Si NPs using the dependence of Optical density on the square of photon energy gives the value of 1.35 eV (Fig, 2,b). This is higher than that of the bulk Si and should be attributed to the quantum size effect.

When the target was irradiated by the Ti:sapphire laser beam with mask, a colloid with a similar absorption spectrum was obtained. The TEM view of NPs is shown in Fig. 3. Subsequent analysis of these NPs using High Resolution TEM and Electron Energy Loss Spectroscopy indicate that they are made of Si with small amount of Si oxide on the surface [4].

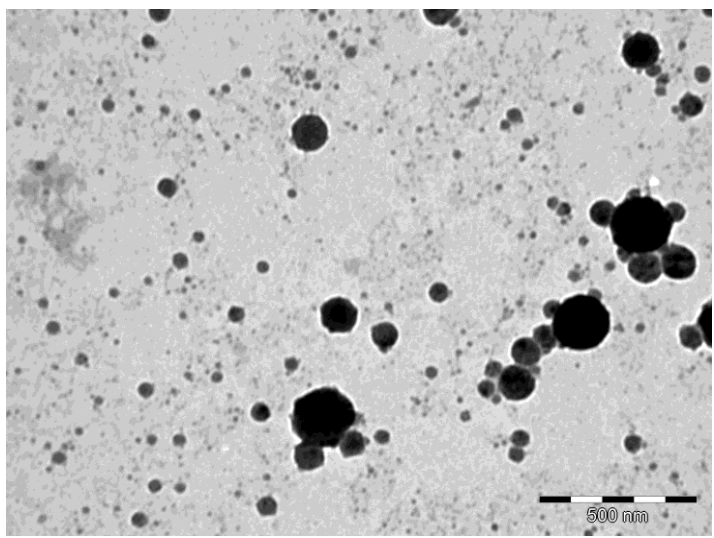

Fig. 3. TEM view of NPs generated by ablation of a bulk Si target with 200 fs pulses of a Ti:sapphire laser. Scale bar denotes 500 nm.

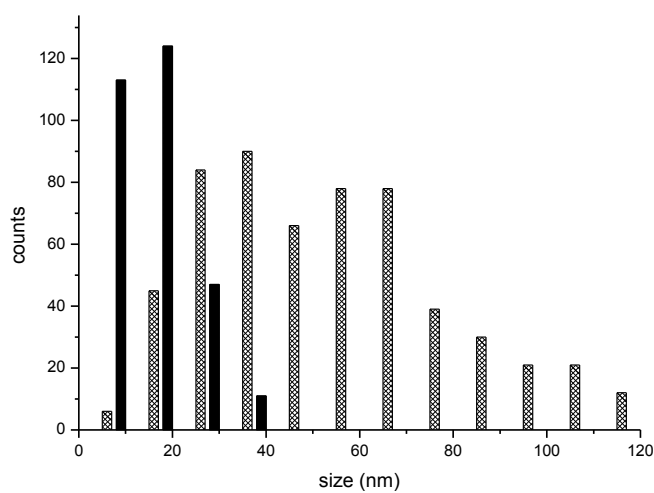

Fig 4. Size distribution of Si NPs generated with a 200 fs pulses in ethanol. Black bars - using mask projection scheme, grey bars - without mask.

The size distribution was determined in the subsequent TEM analysis of the NPs (Fig. 3). One can see that the use of mask shifts the average size of Si NPs toward lower sizes.

3.2 ZnSe nanoparticles

Ablation of ZnSe single crystal was carried out in ethanol using a Cu vapor laser. The colloidal solutions of corresponding NPs were generated either with the mask projection scheme or with a free beam having flat-top profile. The TEM image of these NPs is presented in Fig. 5.

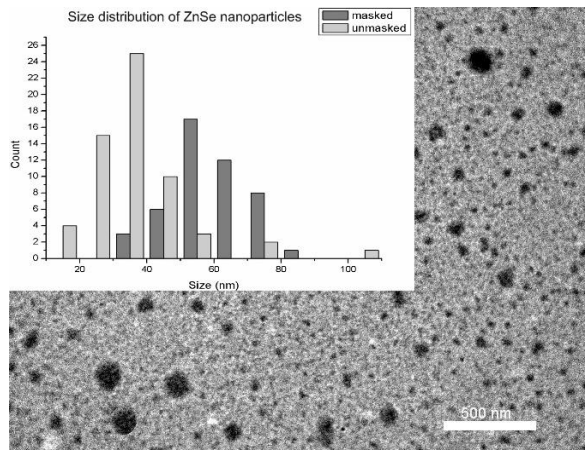

Fig 5. TEM view of NPs of ZnSe generated by ablation of a single crystal ZnSe in ethanol using a Cu vapor laser. The inset shows the size distribution of NPs with and without mask. Scale bar denotes 500 nm.

Unlike ablation of Si, the average size of NPs generated with the mask projection scheme is higher than that without mask. The solutions obtained in both cases are quite similar from the view point of their absorption spectra. The difference lies in the optical properties of NPs. Photoluminescence (PL) spectra of ZnSe colloids are shown in Fig. 6.

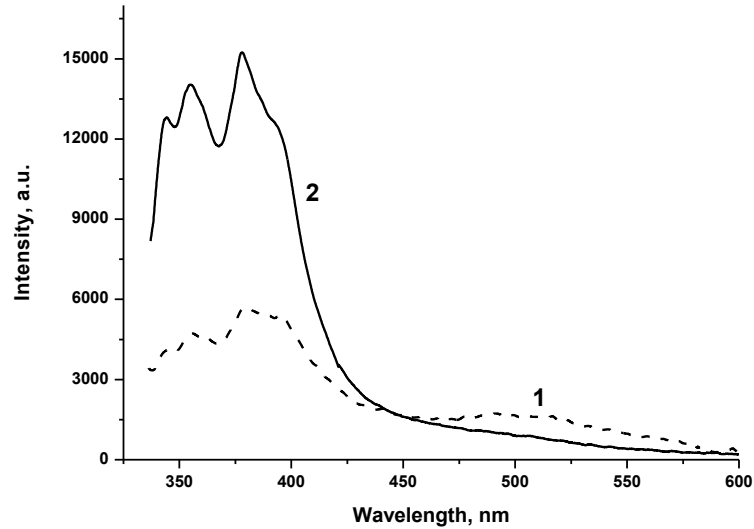

Fig 6. Photouminescence spectra of ZnSe quantum dots produced by ablation of the single crystal ZnSe in ethanol with a Cu vapor laser radiation. Excitation was carried out with a help of a $N_2$-laser at wavelength of 337 nm. 1 - using the mask projection scheme, 2 - without the mask.

The NPs obtained without the mask show no PL in the visible. On the contrary, ZnSe NPs prepared with the mask projection scheme shows a visible-range orange PL. This spectrum is close to that of the PL spectrum of the ZnSe target. It is worth noting that previous experiments of generation of ZnSe quantum dots by ablation of a single crystal ZnSe target in ethanol using unmasked beam of a Cu vapor laser did not result in luminescent NPs [5]. The generated particles were identified as ZnSe according to X-ray diffraction data and Raman analysis. However, those ZnSe NPs showed no PL under excitation in similar experimental conditions.

4. Discussion

Spatial modulation of the beam intensity on the target $I(x,y)$ results in the periodic modulation of the target temperature having the same spatial scale $T(x,y)$. The melt depth also becomes a periodic function of the lateral coordinates, and instead of a continuous molten layer that is formed with a laser beam with smooth profile, a number of smaller melt baths are formed at the same time (see Fig. 7).

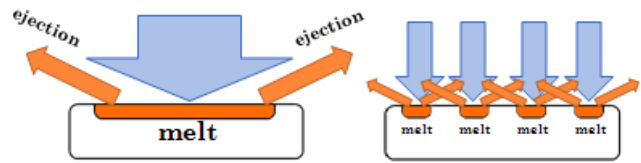

Fig. 7. Sketch of the melt dispersion in case of a laser beam with smooth intensity profile (left) and that of with a laser beam with periodic intensity distribution (right). Blue arrows designate the recoil pressure of vapors of surrounding liquid.

The part of the target within the laser spot is molten during laser exposure. The heat is then transferred to the adjacent layer of the liquid that surrounds the target. As soon as the shock wave leaves this layer, the formation of a gas pocket above the molten layer begins. The target material ejected on the edges of the melt bath is rapidly quenched, since it enters relatively cold liquid. The melt ejected in the center of the laser beam is dispersed into a vapor pocket that is formed above the target. The cooling rate of the ejected nano-entities is smaller than in case of the material ejected on the beam periphery. Moreover, the nano-entities ejected into the vapor can coagulate between each other while they are liquid. Lower cooling rates of the ejected material in the center of the laser beam may also be responsible for the degradation of the target material due to prolonged interaction of the molten material with the vapors of the liquid. In some cases, e.g., for ZnSe quantum dots, this leads to degradation of the properties of resulting NPs, such as their PL.

5. Conclusions

Si nanoparticles obtained in mask projection scheme have lower average size comparing to nanoparticles obtained without the mask. ZnSe quantum dots prepared with the mask projection scheme demonstrate photoluminescence in the visible range of spectrum unlike those obtained without mask. Intensity profile of the laser beam in the experiments on laser-assisted generation of NPs is a crucial parameter that significantly affects their properties, such as size, shape, and their photoluminescence. The obtained results allow revising the model of nanoparticles formation under laser ablation of solids in liquids. Hydrodynamic aspects of NPs formation have been under-estimated in favor of gas-phase condensation of NPs in a plasma plume.

Acknowledgements
The work was partially supported by the Russian Foundation for Basic Research, grants ## 07-02-00757, 08-07-91950, and by Scientific School 8108.2006.2. S.V. Garnov and V.V. Bukin are thanked for their help in work with Ti:sapphire laser.